\documentclass{sig-alternate}
\usepackage{algorithm}
\usepackage{algorithmic}
\usepackage{mathptmx}
\usepackage{graphicx}
\usepackage{amsfonts}
\usepackage{mathptmx}
\usepackage{multicol}
\usepackage{subfloat}
\usepackage{pgfplots}
\usepackage{enumitem}
\usepackage{amssymb}

\usepackage[caption=false]{subfig}
\usepackage{amsmath}

\begin{document}

\title{Sputnik: Elitist Artifical Mutation Hyper-heuristic for Runtime Usage of Multi-objective Evolutionary Algorithms}

\numberofauthors{2} 
\author{
\alignauthor Donia El Kateb and Fran\c{c}ois Fouquet and Yves Le Traon\\
\affaddr{SnT - University of Luxembourg}\\
\email{first.last@uni.lu}
\alignauthor Johann Bourcier\\
\affaddr{University of Rennes 1 / IRISA / INRIA}\\
\email{johann.bourcier@inria.fr}\\ 
}

\maketitle

\begin{abstract}
In the last years, multi-objective evolutionary algorithms (MOEA) have been applied to different software engineering problems where many conflicting objectives have to be optimized simultaneously. 
In theory, evolutionary algorithms feature a nice property for run-time optimization as they can provide a solution in any execution time.
In practice, based on a Darwinian inspired natural selection, these evolutionary algorithms produce many deadborn solutions whose computation results in a computational resources wastage: natural selection is naturally slow. 
In this paper, we reconsider this founding analogy to accelerate convergence of MOEA, by looking at modern biology studies: artificial selection has been used to achieve an anticipated specific purpose instead of only relying on crossover and natural selection (i.e., Muller \textit{et al}~\cite{plaskett1927artificial} research on artificial mutation of fruits with X-Ray). 
Putting aside the analogy with natural selection , the present paper proposes an hyper-heuristic for MOEA algorithms named~\textit{Sputnik}~\footnote{https://en.wikipedia.org/wiki/Sputnik\_virophage} that uses artificial selective mutation to improve the convergence speed of MOEA. 
\textit{Sputnik} leverages the past history of mutation efficiency to select the most relevant mutations to perform.
We evaluate \textit{Sputnik} on a cloud-reasoning engine, which drives on-demand provisioning while considering conflicting performance and cost objectives. 
We have conducted experiments to highlight the significant performance improvement of Sputnik in terms of resolution time.
\end{abstract}

\keywords{MOEA, Hyper-heuristic, Artificial mutation, Optimization, Cloud}


\section{Introduction}
\label{sec:introduction}

Beyond well known harmful computer viruses, Computer Science has also often been inspired by biological processes for more constructive purposes. 
For example, John Holland was inspired by the Darwinian evolution~\cite{darwin1962origin} to design genetic algorithms~\cite{holland1975adaptation,goldberggenetic}. 
These algorithms are particularly efficient to solve complex problems which have very large solution spaces. 
Since their introduction in 1974, they have been applied in numerous contexts (biology, computer science, mathematics) for simulation, optimal allocation, function optimisation.
Multiple-objective evolutionary optimization (MOEA)~\cite{deb2001multi,van2000multiobjective,deb2001multi} is thus a domain that tackles a category of problems in which a decision maker aims at finding a solution that optimizes several conflicting objectives. 
MOEA are used nowadays in several design case studies~\cite{Frey:2013:SGO:2486788.2486856} like cloud organization optimization, however their applicability at run-time presents several issues related to performance.
MOEA algorithms leverage an evolution principle based on Darwinian~\cite{reeves2002genetic} rules that derives, from an initial population of solutions, new solutions combining acceptable trade-offs between objectives. 
Therefore, MOEA algorithms mimic the natural selection by leveraging a set of fitness functions to evaluate a solution to a specific problem.
Additionally, they combine a set of operators~\cite{VanVeldhuizen:1999:MEA:929368} to drive population evolution by introducing small changes on individuals (i.e, mutation), composing new ones (i.e, crossover) or selecting the candidates that will compose next generations (i.e, selection).
However, natural  selection is based on a random mutation in order to mimic the equity of species expressed in Darwinian rules. 
The result of this algorithm is the genotype that is the closest to the optimal solution for the problem.
However, this randomness leads to sub-optimal performance for multiple-objective resolution mainly because of the creation of several unused generations wasting computational resources. 
Besides, modern genetic does not try to mimic the real evolution process in labs. Instead, based on the founding work of Muller~\textit{et al}~\cite{plaskett1927artificial}, artificial mutation is now widely used to save time and generation cycles for instance to produce genetically modified organisms.
Instead of relying only on crossover and natural selection, Muller~\textit{et al}~\cite{plaskett1927artificial} studied artificial mutation using X-Ray to modify a fruit with an anticipated intent. 
These principles have led the genetic field to build instruments for such selective artificial mutation.
Going along the same line, we study how such principles from the genetics could be adapted to MOEA to accelerate the convergence to a solution by guiding the evolutionary algorithms through dynamically selected mutation operators.

Our intuition is that operators applied in a smart and artificial way would provide better results than operators applied randomly, and in particular would reduce the number of useless solutions construction. So, the new algorithm we propose is no longer inspired by Darwinian evolution, but by "artificial mutation", based on a smart and dynamic selection of the best mutation operator to apply at a given step. By applying such operators in priority, we aim at orienting the evolution process of a given population in the right direction for the problem to solve. This makes our new algorithm a new category of hyper-heuristic~\cite{burke2010classification} in the sense it aims at improving performance for a specific problem~\cite{seah2012pareto} as the search evolves. 

In this paper, we present this new hyper-heuristic algorithm, called "Sputnik", inspired by artificial mutation.  Its name "Sputnik" from a virus family which evolves and mutates together with their host in order to perfectly fit their environment and to replicate more quickly.
In the same manner, Sputnik algorithm leverages a continuous ranking of operators according to their impact on fitness functions to smartly select dynamically the mutation operators according to the problem to solve.
We focus on performance as a key factor for runtime usage to reach faster acceptable trade-offs while saving computation time and generation cycles. For instance, the acceleration Sputnik provides is useful for adaptive systems when a solution/reaction has to be found in a short time.
We evaluate our approach on a cloud reasoning engine which handles several conflicting objectives (i.e, latency, cost) and is able to continuously provision customers software by different available cloud providers.
The validation focuses on the performance of our hyper-heuristic for multi-objective resolution.
We have integrated Sputnik in the Polymer framework~\footnote{http://http://kevoree.org/polymer} and evaluated it using a model@runtime platform~\footnote{http://kevoree.org}.
We have conducted an experiment to compare natural selection versus Sputnik. 
Our experiments highlight that Sputnik results in a faster convergence by reducing the number of generations while conserving the ability to achieve acceptable trade-offs on the considered use case.
This paper is organized as follows. 
Section 2 describes the key concepts related to this paper.
Section 3 presents Sputnik hyper-heuristic. 
Section 4 presents validation elements of our approach. 
Finally, section 5 and 6 discuss the related work, our conclusion and future work.

\section{MOEA for at-runtime usage}
\label{sec:background}
Multi-objective evolutionary algorithms (MOEA) are optimization algorithms driven by elitism rules that favor the survival of strongest species~\cite{VanVeldhuizen:1999:MEA:929368} in analogy to natural selection. 
Applied to software engineering, species are candidate solutions to optimization problems which constitute what is called a \textit{generation}. MOEA are based on an iterative search in which a set of individuals is selected and mutated in each iteration to constitute a new \textit{generation}.

\textit{Fitness functions} are used to evaluate solutions to solve a specific optimization problem, in analogy 
to natural selection where species qualities are evaluated according to their surrounding context. 
Genetic operators: \textit{crossover} and \textit{mutation}~\cite{VanVeldhuizen:1999:MEA:929368} are used to emulate natural mutation in order to generate new \textit{generations} composed of mutated solutions.
Mutation operators introduce small changes with a small probability to preserve the population 
diversity~\cite{mahfoud1995niching} whereas crossover operators create new individuals by genetic recombination.
MOEA algorithms can be used to solve at runtime optimization problems, for instance a load balancing problem, therefore performance becomes a major issue to consider~\cite{Jensen:2003:RRC:2221368.2221656}. 
Among the contributions that have addressed these issues, hyper-heuristics~\cite{burke2010classification} are self-tuning heuristics that are based on parameters adaptation during optimization processing. 
In general, hyper-heuristics introduce modifications on the algorithm itself, in order to improve its efficiency to handle a specific purpose~\cite{seah2012pareto}.

The contribution of this paper, detailed in section~\ref{sec:operator}, falls into the category of hyper-heuristics that embed learning mechanisms to achieve performance improvement over a set of software engineering problems~\cite{DBLP:conf/iscis/GuneyKO13,ozcan2008comprehensive}. It relies on an heuristic selection mechanism that evaluates search parameters during the search process and grants higher priority for the most effective ones.

\section{Sputnik}
\label{sec:operator}
\begin{figure}[t!]
  \caption{Sputnik Workflow}
    \label{MOEAWorkflow}
  \centering
    \includegraphics[scale=0.28]{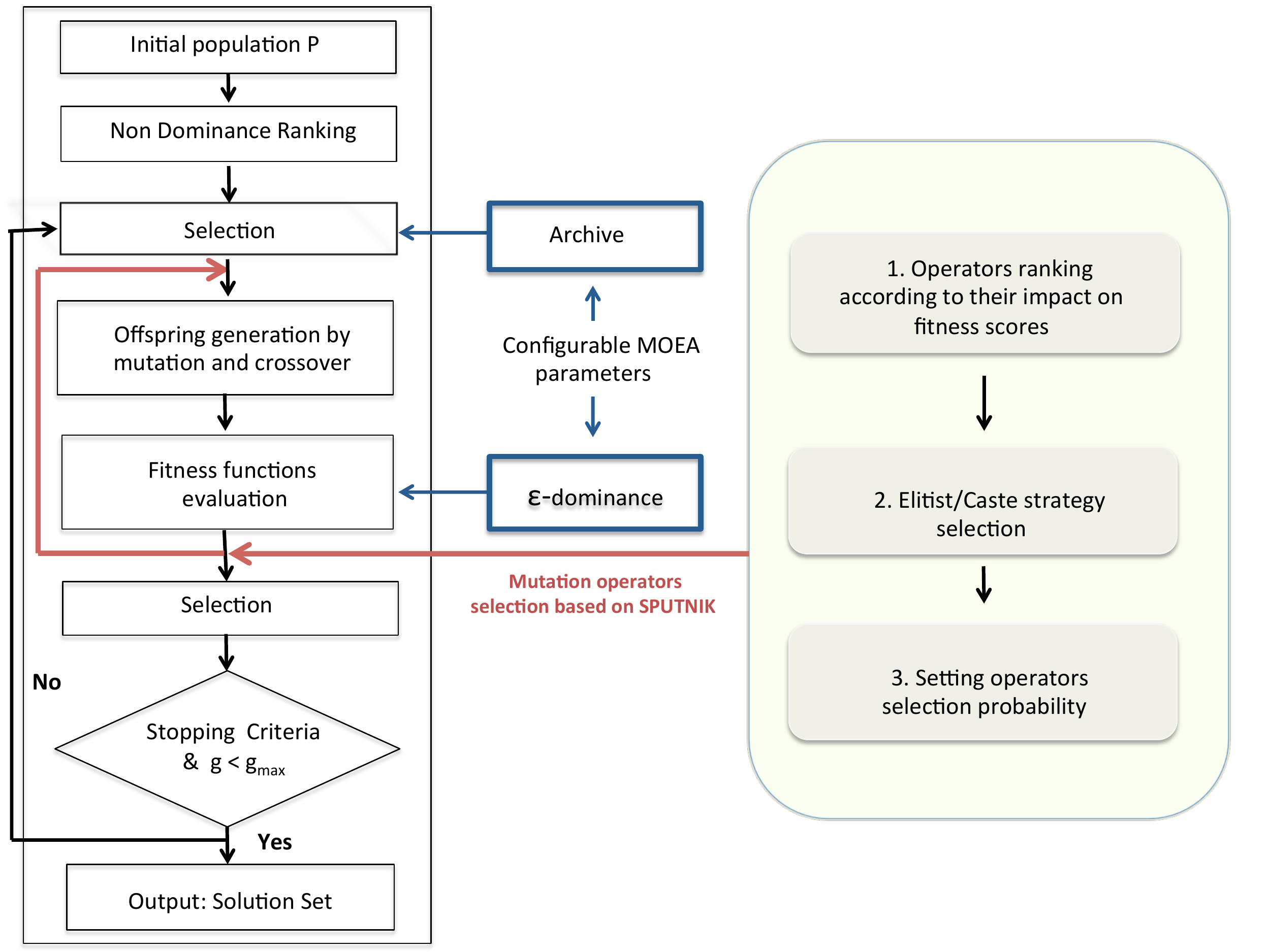}
\end{figure}
\begin{algorithm}[ht!]
\scriptsize
\caption{Sputnik}
\label{Sputnik}
\begin{algorithmic}
\STATE  <Sputnik (Population-Size int, Nb Generation int)>
  \FORALL {j where j range from 1 to Generation-Number} 
 \STATE  Evaluate $f_{i}(x)$ on P new and calculate crowding distance  
\STATE  Update P   
\STATE Apply randomly a set of mutation operators \textit{O} from $O_{set}$ on P to get $P_{new}$ 
\STATE operator-used:=operator-used $\cup$ \textit{O}
\IF{operator-used $\subseteq$ $O_{set}$ } \STATE{evaluate $\triangle_{impact}f_{g_{i}}$ $\forall$ $\textit{O}$ in $O_{set}$ } \ENDIF
\STATE select P $\in$ $\lbrace {P_{elitist}, P_{caste}} \rbrace$
\STATE select \textit{O} from $O_{set}$ with P to get $P_{new}$ with P
\ENDFOR
\end{algorithmic}
\end{algorithm}

The randomness introduced by natural selection of evolutionary approaches leads to suboptimal performance in terms of computational power and memory usage.
Random selection of mutation operators produces useless candidate solutions that lead to computational resources waste and therefore does not meet run-time optimization constraints.
In modern biological studies, after identifying a gene impact on individual phenotype trait, scientists like Muller~\textit{et al}~\cite{plaskett1927artificial} leverage artificial mutation to directly produce an individual combining the foreseen modification.

Our hypothesis is that the artificial mutation concept can be introduced in evolutionary algorithms to mimic modern biological genetic, thus reducing the number of required generations to reach acceptable solutions.
The \textit{a priori} scientific knowledge of a gene modification impact, could be replaced by a continuous ranking and learning approach leveraging execution history.
Therefore, in this paper we aim at optimizing MOEA, by dynamically reducing the usage of mutation operators that are less effective in improving fitness functions scores. 
At the same time, we maintain the equity of natural selection, to ensure that the modified evolution algorithm is able to reach any solution.
Thus, we replace the random mutation operator selection by an hyper-heuristic that detects for each individual the most pertinent operator to apply in order to achieve a faster trade-off. 

After each application, mutation operators are classified according to the delta variance they introduce on each fitness function.
Internally, Sputnik maintains an elitist group of mutation operators that are relevant to improve a fitness function score.
To enhance operators selection, Sputnik, considers the current fitness scores reached by a solution, and selects the most relevant mutator in elite groups to improve next generation~\footnote{http://www.genetics.org/content/111/1/147.short}.

Sputnik workflow, depicted by the flowchart of Figure~\ref{MOEAWorkflow}, takes as inputs an initial population and a generation number. As most of MOEA variants like (NSGA-II, $\epsilon$-MOEA, SPEA 2~\cite{van2000multiobjective}) the major part of the algorithm is an individual ranking step according to fitness function evaluation and a non dominating population construction.
Sputnik introduces a favoritism operator approach in the mutation process. 

We consider a multi-objective evolutionary optimization of f with n objectives $(f_{1}$, $f_{2}$,...,$f_{n})$.
The average fitness score for a generation $g_{i}$ is defined by $\sum_{i=1}^n f_{i}/n$.
We define $\triangle_{impact}f_{g_{i}}$ as the fitness score variation between the average of fitness function evaluation for a generation $g_{i-1}$ and a generation $g_{i}$: $\triangle_{impact}f_{g_{i}}$=$(\sum_{i=1}^n f_{i}/n)_{g_{i}}$-$(\sum_{i=1}^n f_{i}/n)_{g_{i-1}}$.
Sputnik records the \textit{selection occurrence} for the different mutation operators that have been involved in each generation.
Once all mutators have been selected at least once, $\triangle_{impact}f_{g_{i}}$ is evaluated for all the operators and Sputnik can be configured to select the operators that have $\triangle_{impact}f_{g_{i}}$ > 0 in the generation $g_{i+1}$ with the two following settings:
\begin{itemize}[label=$\bullet$,leftmargin=* ,parsep=0cm,itemsep=0cm,topsep=0cm]
\item \textit{Elitist Strategy}: The operator that has the highest \textit{delta impact positive} is always chosen in the generation $g_{i+1}$. This configuration accords higher chance to the "winner operator" to be selected in the next generation.
\item \textit{Caste Strategy}: A selection probability is partitioned between operators that have 
$\triangle_{impact}f_{g_{i}}$ > 0 and is defined as \textit{$P_{selection}$} = 
$\triangle_{impact}f_{g_{i}} \diagup \sum_{i=1}^n \triangle_{impact}f_{g_{i}}$
This configuration gives more equity in terms of selection probability for all operators which have a positive impact on a fitness.
\end{itemize}
SPUTNIK-based mutation operators selection is described in Algorithm 1.
In both settings a probability of 10\% is maintained for pure random selection of operators to not discriminate worst ranked operators. 
Sputnik keeps 10\% of mutation to give chance to less selected operators to be reintroduced in the elite group of a fitness function. 
By this mechanism we keep a minimal equity of species while conserving 90\% of the Pareto for most efficient mutation.
\vspace{-0.5cm}
\section{Validation}
\label{sec:validation}
We validate our approach through a reasoning engine for an hybrid cloud management system:
A cloud customer aims at hosting some of his applications on a set of virtual machines that are dedicated for his internal usage in a private cloud provider. The customer aims at reducing costs inherent in a setting based on private cloud model by exporting some applications to a public cloud. The problem can be formulated as follows: ``Given a dedicated set of VMs allocated by a private cloud provider, how to optimize software placement in VMs to reduce \textbf{costs} and to reduce \textbf{latency} that can be introduced by a distant hosting?".

We define a cloud configuration as an architecture model which leverages virtual machine and component (i.e, provisioned software in our context) concepts. Based on our architectural model, an \textit{Individual} represents a solution vector $x\in X$ that corresponds to a cloud infrastructure model. 
A \textit{gene} corresponds to a component or a virtual machine in our model. 
A \textit{population} corresponds to a set of cloud infrastructure models.
A \textit{Genetic mutation operator} corresponds to an elementary flip in the model like \textit{AddComponent(c,A)}. 
For the sake of space limitations, we will not give ample details about the operators defined in this paper. The reader may refer to our open source Polymer framework which gives ample details about our architectural 
model~\footnote{https://github.com/dukeboard/kevoree-genetic}.

A Cloud infrastructure multi-objective optimization problem is represented by the following Triplet \textit{(I,F,CO)}. 
{I} denotes a cloud infrastructure model which represents an abstraction of 
a set of (VM). Each (VM) hosts n software components (C).
\textit{CO} denotes a set of possible configurations in \textit{I} that satisfy F. A configuration $co \in CO$ is obtained through a mapping from components to (VMs), for example $co=(VM_{1}(c_{1},c_{2},c_{3}))$ denotes a configuration with a single virtual machine $VM_{1}$ hosting 3 components $c_{1},c_{2},c_{3}$. 
The vector \textit{F}(X) is composed of the following 2 objective functions, 
$F(X)=(f_{1}(x),f_{2}(x))$ that have to be minimized:
1) $f_{1}(x)$= Cost(x): denotes the cost which is proportional to the number of active VMs.
2) $f_{2}(x)$= Latency(x): The latency function is calculated on the basis of the average of latency of all components in the model, considering both redundancy and distant hosting impacts.

The multi-objective optimization problem aims at finding a configuration $co \in CO$ such as min $\underset{co}{F(X)}$.
We have implemented Sputnik prototype in the Polymer framework which leverages a model based encoding to perform MOEA optimization.
Indeed, using model@run.time paradigm~\cite{5280648}, our cloud models can be seamlessly deployed in a real large-scale production environment~\cite{DBLP:conf/dais/FouquetDPBBJ12} like a cloud infrastructure. 
The evaluation of cloud Sputnik based reasoning engine is out of scope of this paper, only the gain achieved by the hyper-heuristic will be evaluated in this paper.
\subsection{Research Questions}

This validation section aims at exploring the effectiveness of Sputnik to reduce the generations number to reach a certain level of trade-off. 
Secondly, we provide evidence that the results are comparable in terms of level of trade-off achieved. 

Finally, we explore the effectiveness of Sputnik once embedded in most popular MOEA algorithms such as MOEA-D and NSGA-II. To compare the solutions of the different algorithms under study, we choose the hypervolume metric as Pareto-front quality indicator~\cite{zitzler2007hypervolume}. Our validation steps can be summarized in the following research questions:
\textbf{$RQ_{1}$}) Considering that an acceptable trade-off is 90\% of the best solutions, is the Darwin Sputnik operator selection strategy successful to reduce the number of generations to reach the acceptable trade-off compared to a classical random strategy?
\textbf{$RQ_{2}$}) Does Sputnik produce comparable results in terms of trade-off achieved regarding classical random mutation selection?
\textbf{$RQ_{3}$}) Is Sputnik relevant for common MOEA algorithms?

\subsection{Experimental results}
To answer $RQ_{1}$, we run Sputnik in a cloud optimization engine which contains 100 virtual nodes and several components that have to be dispatched. 
Our cloud reasoning engine is configured with one of the most popular MOEA algorithm NSGAII~\cite{deb2002fast}.
The experiment is run 3 times for 300 generations, Sputnik with caste strategy, Sputnik with elitist strategy and finally with random. The results are shown in the Figure 2.
We consider that a solution achieves an acceptable trade-off if it reaches 90\% of the best obtained solution. 
In our case study, the best obtained solution achieves an hyper-volume of 0.79 (acceptable hypervolume value is 0.71 in our case), it has been reached with a 2000 generations previous run.
A run with Sputnik (with both strategies) reaches this value after 176 generations whereas a run with random operators selector reaches this value after 279 generations.
Sputnik strategies are very similar in terms of hypervolume achievements.
We notice that elitist converges slighly faster however, the caste strategy can reach better hyper-volume scores because it takes benefit from mutators diversity.
In average, Sputnik (both with cast and elitist) outperforms random selector by reducing around 37\% the number of necessary generations. 
This confirms our first hypothesis which states that a smart mutators selection strategy is successful to reduce the number of generations to reach an acceptable trade-off compared to a classical random selection strategy.

To reply to $RQ_{2}$, we run a similar experiment with both random and Sputnik selector until we reach an unchanged value of hypervolume for 50 generations. 
Final values for Sputnik (elitist: 0.77 and caste: 0.81) and random (0.78) ($\pm$1\% each difference) allow us to conclude that our hyper-heuristic does not decrease the quality of the result in terms of degree of trade-off achieved. 
Moreover the caste strategy improves the hypervolume score.

For $RQ_{3}$, we set up common MOEA algorithms with Sputnik.
The results shown in Table~\ref{tableresult1}, are hypervolume values after 200 generations compared to MOEA without Sputnik. 
We conclude that Sputnik (both with caste and elitist configurations) is effective in achieving a better trade-off faster than current MOEA.
\begin{figure}
\centering
\begin{tikzpicture}[line width=0.3pt,scale=0.96]
\scriptsize
\begin{axis}[legend entries={SPUTNIK (ELISTIST),SPUTNIK (CASTE), RANDOM},
reverse legend, legend style={at={(0.85,0.3)}}, xlabel=Generation Number, ylabel=Hypervolume]
\addplot +[only marks,mark=o] table [x=generation, y=SPUTNIK_ELITIST,  width=.8pt, col sep=comma] {result2.csv};
\addplot +[only marks] table [x=generation, y=SPUTNIK_CASTE, col sep=comma] {result1.csv};
\addplot +[only marks,mark=x] table [x=generation, y=RANDOM, col sep=comma] {result1.csv};
\end{axis}
\end{tikzpicture}
\label{MOEAresults}
\caption{Hypervolume: Sputnik (Elitist \& Caste) vs Random }
\vspace{-0.30cm}
\end{figure}
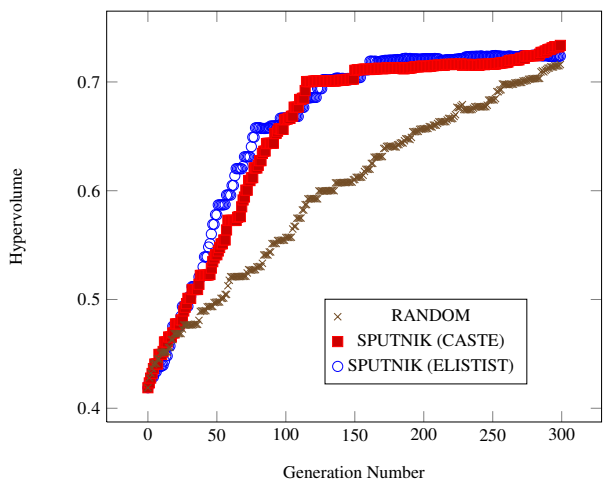

\begin{table}[h!]
\begin{center}
\scriptsize
  \begin{tabular}{| c | c | c |  c | c | }
    \hline
   \textbf{}  & \textbf{NSGAII} & \textbf{$\epsilon$-NSGA II } & \textbf{$\epsilon$-MOEA} & \textbf{Hyper-MOEA} \\\hline
     \textbf{Elitist}  & \textbf{0.72} & \textbf{	0.75} & \textbf{0.51} & \textbf{0.47} \\ \hline
     \textbf{Caste}  & \textbf{0.70} & \textbf{0.73} & \textbf{0.52} & \textbf{0.47} \\ \hline
       \textbf{Random }  & \textbf{0.67} & \textbf{0.66 } & \textbf{0.43} &\textbf{0.44}  \\ \hline  		
  \end{tabular}
    \caption{Sputnik (Elitist \& Caste) hypervolume on MOEA}    
    \label{tableresult1}
    
\end{center}
\end{table}
\vspace{-0.99cm}
\section{Related Work}
\label{sec:relatedwork}
Several approaches have proposed hyper-heuristics that operate on top of MOEA to improve their run-time usage.
In~\cite{seah2012pareto}, the authors embed learning techniques in classical MOEA. 
They assume that objective functions are expensive to compute so they rank the pareto front elements and they evaluate only the individuals that have higher ranks.
In~\cite{leon2009hyperheuristics}, the authors have proposed an hyper-heuristic that relies on the hypervolume calculation to improve computational results. These approaches consider only the Pareto front set evaluation to improve MOEA, whereas our approach evaluates operators contribution in improving fitness and thus injects mutation operators that are eligible to make MOEA converge faster. 
In~\cite{Yuan04statisticalracing}, the authors have shown that racing algorithms can be used to reduce the computational resources inherent from using evolutionary algorithms in large scale experimental studies, their approach automates solutions selection and discards solutions that can not come up with results improvement.
Whereas racing techniques eliminate worst solutions candidates to speed up the search, in our approach 
we keep considering worst ranked candidates to maintain operators diversity.
In~\cite{drake2011controlling}, the authors have explored the advantages of using a controlled crossover on top of 
single-point search based hyper-heuristics. 
They maintain the best solutions obtained during the search and update crossover operator accordingly. 
The authors rely on a process focused on crossover as a biological selective breeding.
This breeding assumes that fittest genes are already present in the initial population.
Unlike cited approaches, Sputnik focuses on artificial mutation selection, therefore mimicking the process used to produce genetically modified organisms.
As far as we know, there is no other hyper-heuristics that proposes artificial mutation at mutators level.

\section{Conclusion}
\label{sec:conclusion}
In this paper, we have introduced a novel hyper-heuristic breaking the random natural selection of classical MOEA to leverage instead an elitist artificial mutation inspired by biological studies~\cite{plaskett1927artificial} supported by a continuous learning of mutations impact.
Sputnik, relies on a mutation operator selection based on a continuous learning of past effect on fitness functions. 
The overall goal of Sputnik is to enhance the optimization algorithm itself, and to guide the search towards faster trade-off achievement to finally save generation cycle and time.
Integrated in a model-based optimization framework, our approach has been validated on the construction of a cloud reasoning engine prototype. 
Experimentally, we provide evidence of the effectiveness of artificial guided mutation to reduce significantly the number of necessary generations to find acceptable trade-offs.
In the future, we plan to evaluate the relevancy of our heuristic on crossover operations. 
Also we plan to classify elitist strategies to analyze their impact on solution diversity.
\vspace{-2.36mm}
\bibliographystyle{abbrv}
\bibliography{sigproc} 

\begin{thebibliography}{10}

\bibitem{5280648}
G.~Blair, N.~Bencomo, and R.~France.
\newblock Models@ run.time.

\bibitem{darwin1962origin}
C.~Darwin and J.~Burrow.
\newblock {\em The Origin of Species by Means of Natural Selection}.
\newblock YNY, 1962.

\bibitem{deb2001multi}
K.~Deb.
\newblock Multi-objective optimization.
\newblock {\em Multi-objective optimization using evolutionary algorithms}.

\bibitem{deb2002fast}
K.~Deb, A.~Pratap, S.~Agarwal, and T.~Meyarivan.
\newblock A fast and elitist multiobjective genetic algorithm: Nsga-ii.
\newblock {\em EC}, 2002.

\bibitem{drake2011controlling}
J.~H. Drake, E.~{\"O}zcan, and E.~K. Burke.
\newblock Controlling crossover in a selection hyper-heuristic framework.
\newblock 2011.

\bibitem{burke2010classification}
B.~et~al.
\newblock A classification of hyper-heuristic approaches.
\newblock In {\em Handbook of Metaheuristics}.

\bibitem{Frey:2013:SGO:2486788.2486856}
F.~et~al.
\newblock Search-based genetic optimization for deployment and reconfiguration
  of software in the cloud.
\newblock In {\em ICSE}, 2013.

\bibitem{DBLP:conf/dais/FouquetDPBBJ12}
F.~F. et~al.
\newblock Dissemination of reconfiguration policies on mesh networks.
\newblock In {\em DAIS}, 2012.

\bibitem{DBLP:conf/iscis/GuneyKO13}
I.~A.~G. et~al.
\newblock Hyper-heuristics for performance optimization of simultaneous
  multithreaded processors.
\newblock In {\em ISCIS}, 2013.

\bibitem{seah2012pareto}
S.~et~al.
\newblock Pareto rank learning in multi-objective evolutionary algorithms.
\newblock In {\em CEC 2012}.

\bibitem{zitzler2007hypervolume}
Z.~et~al.
\newblock The hypervolume indicator revisited.
\newblock In {\em EMO}, 2007.

\bibitem{goldberggenetic}
D.~Goldberg.
\newblock Genetic algorithms in search, optimization, and machine learning,
  addison-wesley, reading, ma, 1989.

\bibitem{holland1975adaptation}
J.~H. Holland.
\newblock {\em Adaptation in natural and artificial systems}.
\newblock U Michigan Press, 1975.

\bibitem{Jensen:2003:RRC:2221368.2221656}
M.~T. Jensen.
\newblock Reducing the run-time complexity of multiobjective eas: The nsga-ii
  and other [...].
\newblock {\em EC}, 2003.

\bibitem{leon2009hyperheuristics}
C.~Le{\'o}n, G.~Miranda, and C.~Segura.
\newblock Hyperheuristics for a dynamic-mapped multi-objective island-based
  model.
\newblock 2009.

\bibitem{mahfoud1995niching}
S.~W. Mahfoud.
\newblock Niching methods for genetic algorithms.
\newblock {\em Urbana}, 1995.

\bibitem{ozcan2008comprehensive}
E.~{\"O}zcan, B.~Bilgin, and E.~E. Korkmaz.
\newblock A comprehensive analysis of hyper-heuristics.
\newblock {\em Intelligent Data Analysis}, 2008.

\bibitem{plaskett1927artificial}
H.~Plaskett.
\newblock Artificial transmutationof the gene.
\newblock {\em TIC 1927}.

\bibitem{reeves2002genetic}
C.~R. Reeves and J.~E. Rowe.
\newblock {\em Genetic algorithms-principles and perspectives}.
\newblock 2002.

\bibitem{VanVeldhuizen:1999:MEA:929368}
D.~A. Van~Veldhuizen.
\newblock {\em Multiobjective evolutionary algorithms: classifications,
  analyses, and new innovations}.
\newblock PhD thesis, 1999.

\bibitem{van2000multiobjective}
D.~A. Van~Veldhuizen and G.~B. Lamont.
\newblock Multiobjective evolutionary algorithms: Analyzing the
  state-of-the-art.
\newblock {\em Evolutionary computation}, 2000.

\bibitem{Yuan04statisticalracing}
B.~Yuan and al.
\newblock Statistical racing techniques for improved empirical evaluation of
  evolutionary algorithms.
\newblock 2004.

\end{thebibliography}
\end{document}